# An Adaptive Intelligent Thermal-Aware Routing Protocol for Wireless Body Area Networks

Abdollah Rahimi[1] . Mehdi Jafari Shahbazzadeh[2,*] . Amid Khatibi[3]

**Abstract:** Wireless Body Area Networks (WBANs) have gained significant attention due to their widespread applications in healthcare monitoring, sports performance tracking, military communication, and remote patient care. These networks consist of wearable or implanted sensor nodes continuously collecting and transmitting physiological data, requiring an efficient and reliable communication framework. However, the unique challenges of WBANs, such as limited energy resources, dynamic network topology, and high sensitivity to node temperature, necessitate specialized routing strategies. Traditional routing protocols, which often prioritize shortest-path selection, tend to create traffic congestion and overheating in specific nodes, leading to early network failures and reduced overall performance. To address these issues, this paper proposes an intelligent, temperature-aware, and reliability-based routing approach that enhances the overall efficiency and stability of WBANs. The proposed method operates in two phases: (1) network setup and intelligent path selection and (2) dynamic traffic management and hotspot avoidance. In the first phase, sensor nodes exchange vital network status information, including residual energy, node temperature, link reliability, and delay, to build an optimized network topology. Instead of relying solely on shortest-path routing, a multi-criteria decision-making algorithm is employed to select the most efficient paths, prioritizing those that balance energy consumption, temperature regulation, and communication stability. This prevents excessive energy depletion in specific nodes and avoids forming traffic bottlenecks. The system continuously monitors real-time network conditions in the second phase, dynamically rerouting traffic away from overheated or energy-depleted nodes. This ensures that critical sensor data is reliably delivered while extending the network's lifetime. Simulation results demonstrate the superiority of the proposed approach compared to existing methods. The proposed method improves throughput by 13% and reduces end-to-end delay by 10%. Additionally, it achieves a 25% reduction in energy consumption. The proposed method also significantly reduces the normalized routing load by 30%.

**Keywords** Internet of Things . Wireless Body Area Networks . Temperature-Aware Routing . Link Quality

✉ Abdollah Rahimi
  a.rahimi.iauk.ac@gmail.com
✉ Mehdi Jafari Shahbazzadeh*
  mjafari@iauk.ac.ir
✉ Amid Khatibi
  a.khatibi@srbiau.ac.ir

1 Department of computer Engineering, Kerman Branch, Islamic Azad University, Kerman, Iran.
2 Department of Electrical Engineering, Kerman Branch, Islamic Azad University, Kerman, Iran.
3 Department of computer Engineering, Bardsir Branch, Islamic Azad University, Bardsir, Iran.

# 1 Introduction

WBANs have emerged as a pivotal technology in modern healthcare, enabling real-time monitoring of patients through wearable and implantable sensors. These networks facilitate continuous health data transmission to medical professionals, enhancing early diagnosis and timely intervention [1]. Figure 1 illustrates a three-layer communication architecture in WBANs designed for healthcare applications. This architecture ensures seamless data transmission from body sensors to medical professionals for real-time monitoring and emergency response. The first layer (Intra-WBAN Communication) consists of multiple body sensors placed on a patient's body to continuously monitor physiological parameters such as heart rate, body temperature, and oxygen levels. These sensors wirelessly transmit data to a central hub, such as a smartphone or a wearable device, which serves as a local processing unit. The second layer (Inter-WBAN Communication) facilitates data transfer from the central hub to nearby access points, computers, or the Internet using Bluetooth, Wi-Fi, or other short-range communication technologies. This layer ensures real-time data relay for further analysis. The third layer (Beyond-WBAN Communication) involves transmitting health data through communication towers, access points, or the Internet to medical servers, databases, and healthcare providers. This layer enables medical professionals, emergency response teams, and healthcare centers to access patient data remotely, ensuring timely diagnosis and intervention.

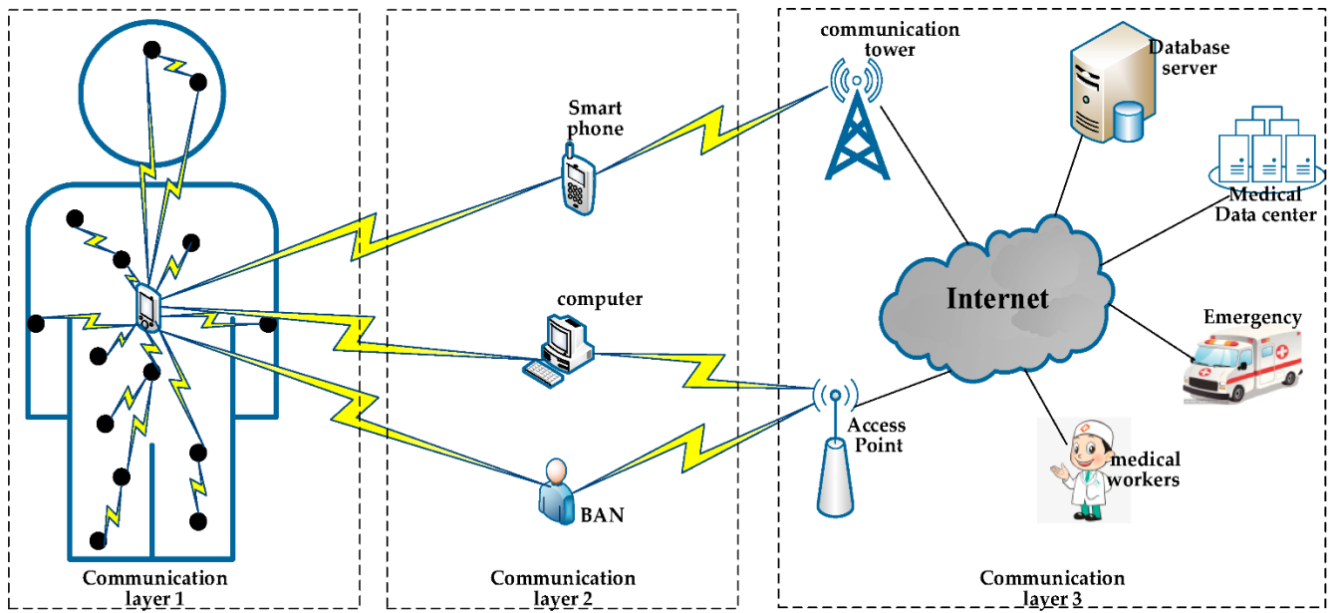

**Fig .1** Architecture of WBAN.

Despite their advantages, WBANs face critical challenges, including energy efficiency, link reliability, and thermal management, all of which impact the longevity and effectiveness of these networks [2]. Addressing these challenges is essential for ensuring seamless and secure data transmission in healthcare applications. One of the most pressing issues in WBANs is energy consumption, as sensor nodes operate on limited battery power, and frequent recharging or replacement is impractical [3]. Energy-efficient routing strategies are crucial to optimizing network lifespan while maintaining data integrity and reliability [4]. Additionally, link reliability is a key factor affecting data delivery, as the human body introduces dynamic channel conditions, frequent link disconnections, and interference that degrade communication quality [5]. Therefore, routing protocols must incorporate link quality assessment to enhance communication stability. Beyond energy and reliability concerns, thermal awareness has gained significant attention in WBANs. Continuous data transmission can cause a rise in temperature in sensor nodes, which may lead to tissue damage and sensor failure if not properly managed [6]. Existing routing protocols primarily focus on energy efficiency and link quality while overlooking the potential risks of heat generation in WBANs [7]. As a result, there is a strong need for a routing strategy that integrates energy efficiency, link reliability, and thermal awareness to ensure network longevity, stable connectivity, and patient safety [8]. Several studies have proposed energy-efficient and multi-path routing techniques to mitigate congestion and reduce communication overhead in WBANs [9]. However, most of these approaches lack an integrated mechanism to handle concurrent energy depletion, temperature rise, and link instability. This paper presents a novel Thermal-Aware and Energy-Efficient Routing Protocol that leverages dynamic path selection based on real-time temperature, energy levels, and link reliability. The proposed method ensures optimal load balancing by diverting traffic away from overheated nodes while maximizing network lifetime and maintaining low latency.

## 1.1 Problem statement

- **Thermal Management**: Continuous data transmission causes sensor nodes to overheat, leading to tissue damage and reduced sensor lifespan [6]. Most existing routing protocols fail to incorporate real-time thermal management or rely on static threshold-based mechanisms, which lack adaptability to dynamic network conditions [7]. To prevent hotspot formation, a dynamic temperature-aware routing strategy is needed to reroute traffic away from overheated nodes [8].
- **Energy Consumption and Network Lifetime**: WBAN sensor nodes have limited energy resources, and replacing or recharging them is challenging, especially for implanted devices [3]. Some existing routing protocols focus solely on minimizing energy consumption, which often leads to overburdening certain nodes, causing premature energy depletion and network failure [4]. A balanced energy-efficient routing mechanism is required to distribute energy consumption evenly among nodes and extend the network lifetime [9].
- **Link Reliability and Communication Stability**: Body movement, interference, and channel variations cause frequent link disconnections and packet loss, leading to delays in data transmission [5]. Traditional routing protocols prioritize shortest paths without considering link stability, which results in frequent retransmissions and reduced communication reliability [6]. A reliability-aware routing strategy must be implemented to select stable links with a high packet reception ratio (PRR) and minimal disconnections [10].

## 1.2 Contributions

Our main contributions are unfolded as follows:

- **Intelligent Thermal-Aware Routing Mechanism:** A real-time temperature monitoring and management strategy is implemented to prevent sensor overheating, ensuring patient safety. The protocol dynamically excludes overheated nodes from routing and redistributes traffic to prevent hotspot formation.
- **Energy-balanced routing with Adaptive Load Distribution:** A multi-criteria optimization algorithm ensures even energy consumption among sensor nodes, preventing premature node failures. The protocol selects paths based on energy availability, reducing the overuse of specific nodes and extending the network lifespan.
- **Reliability-Aware Routing with Dynamic Link Selection:** A packet reception ratio (PRR)-based link assessment ensures that only stable and high-quality links are used for routing. The method accounts for body movement, interference, and environmental changes, reducing packet loss and retransmissions. An adaptive path selection mechanism continuously updates routes to maintain stable communication, reducing latency and improving Quality of Service (QoS).
- **Traffic Management and Congestion Avoidance Strategy:** A real-time congestion detection mechanism prevents excessive packet buildup and optimizes network performance. Critical health data is prioritized by introducing priority-based data transmission, ensuring minimal delay for emergency information.

## 1.3 Paper Organization

The remaining sections of this paper are organized as follows: Section II describes the related work. Section III introduces the proposed method. Section IV presents the experimental results. Finally, Section V concludes the paper. Table 1 shows the abbreviations and their full forms.

**Table 1.** Abbreviations.

| Abbreviation | Full Form | Abbreviation | Full Form |
|---|---|---|---|
| WBAN | Wireless Body Area Network | PRR | Packet Reception Ratio |
| IoT | Internet of Things | SAR | Specific Absorption Rate |
| MAC | Medium Access Control | TDMA | Time Division Multiple Access |
| QoS | Quality of Service | CBR | Constant Bit Rate |
| EWMA | Exponentially Weighted Moving Average | NRL | Normalized Routing Load |
| AODV | Ad hoc On-Demand Distance Vector | AI | Artificial Intelligence |

# 2 Related Work

In [11], Khan et al. proposed RLT, a reliable, link quality, and temperature-aware routing protocol for WBANs. Their approach integrates temperature monitoring and link quality assessment to enhance communication reliability and prevent excessive heating in sensor nodes. Experimental results demonstrated that RLT reduces packet loss and maintains stable network performance while effectively mitigating temperature-related risks. A study by Singla et al. [12] introduced an optimized energy-efficient and secure routing protocol for WBANs. Their method combines energy-awareness with security measures, ensuring reliable data transmission while protecting against potential cyber threats. Findings indicate that the proposed approach enhances energy efficiency, extends network lifespan, and improves data security compared to conventional routing techniques. In [13], Ullah et al. presented an energy-efficient and reliable routing scheme to enhance network stability in WBANs. Their protocol dynamically adjusts routing paths based on residual energy levels and link stability, preventing premature node failures. Simulation results revealed that the scheme significantly improves network stability, reduces energy consumption, and extends the operational lifespan of WBANs.

Zaman et al. [14] proposed EEDLABA, an energy-efficient, distance-aware, and link-aware body area routing protocol. Their approach incorporates a clustering mechanism to optimize energy utilization and link reliability while maintaining stable network performance. Performance evaluations showed that EEDLABA outperforms traditional routing protocols regarding energy conservation, reduced latency, and improved packet delivery ratio. A study by Aryai et al. [15] introduced SIMOF, a swarm intelligence multi-objective fuzzy thermal-aware routing protocol for WBANs. This protocol applies fuzzy logic and swarm intelligence techniques to dynamically balance energy consumption and temperature regulation. Results demonstrated that SIMOF efficiently prevents hotspot formation, extends network lifetime, and maintains reliable communication across WBAN nodes. Ahmed et al. [16] developed a thermal and energy-aware routing protocol to optimize heat distribution and energy consumption in WBANs. Their method leverages adaptive path selection to reroute data away from overheated nodes while ensuring energy-efficient transmissions. Findings indicate that this approach reduces thermal stress on sensors, improves energy efficiency, and enhances overall network reliability.

In [17], Raed and Alabady proposed a hotspot-aware, multi-cost-based energy-efficient routing protocol for WBANs. Their methodology considers temperature variations, energy levels, and communication costs to dynamically select optimal routing paths. Simulation results showed that their protocol successfully reduces hotspots, balances energy consumption, and maintains a stable network connection. Kim et al. [18] presented an enhanced temperature-aware routing protocol for WBANs. Their technique employs real-time temperature monitoring and dynamic load balancing to prevent overheating in sensor nodes. The study demonstrated that the protocol effectively minimizes sensor temperature spikes and enhances data transmission reliability. Abdullah [19] proposed an energy-aware and reliable routing protocol to improve WBAN performance. Their approach prioritizes nodes with stable energy levels and reliable links, ensuring efficient data forwarding. Findings indicate that this protocol enhances network lifetime, reduces communication delays, and optimizes energy utilization across WBAN nodes.

In another study [20], Zuhra et al. presented MIQoS-RP, a multi-constraint intra-BAN QoS-aware routing protocol for wireless body sensor networks. This protocol considers multiple quality of service factors, such as latency, reliability, and energy efficiency, to enhance network performance. Findings indicate that MIQoS-RP outperforms traditional routing protocols in ensuring QoS compliance and prolonging network lifespan. [21] introduced ZEQoS, a new energy and QoS-aware routing protocol designed for sensor device communication in healthcare systems. Their protocol balances energy consumption while ensuring reliable data transmission. Results indicate that ZEQoS improves network performance by reducing energy depletion and enhancing data delivery rates. In [22], Al_Barazanchi et al. proposed a novel routing protocol to reduce path loss in WBANs. Their approach optimizes communication paths to minimize signal attenuation and improve data reliability. The study demonstrated that their method significantly enhances signal strength and network efficiency.

Ahmad et al. developed an energy-efficient framework for WBANs in the healthcare domain, integrating intelligent energy management techniques to extend network lifetime. Their framework dynamically adjusts transmission power based on network conditions to optimize performance. Findings show improved energy conservation and prolonged node lifespan compared to conventional approaches [23]. According to [24], a link quality and energy utilization-based routing strategy was proposed for WBANs. Their method selects the next-hop node based on energy levels and link stability, ensuring efficient data transmission. Simulation results highlighted significant network reliability and efficiency improvements over existing protocols. In another study [25], Bedi et al. presented a novel routing protocol leveraging Grey Wolf Optimization and Q-learning for WBANs. This hybrid approach optimizes routing paths by dynamically adapting to network conditions. Experimental evaluations confirmed that their protocol enhances data transmission reliability and reduces energy consumption.

Roopali and Kumar proposed an energy-efficient dynamic cluster head and routing path selection strategy for WBANs. Their method employs adaptive clustering to balance energy consumption across sensor nodes. Results demonstrated that this approach extends network lifespan and reduces communication overhead compared to conventional clustering techniques [26]. In [27], Elhadj et al. introduced a priority-based cross-layer routing protocol for healthcare applications. Their protocol prioritizes medical data based on urgency, ensuring timely and reliable delivery. Findings indicate that their approach enhances the quality of service and minimizes transmission delays in critical healthcare scenarios.

Recent research has also explored intelligent and energy-aware routing in renewable sensor networks. Aslam et al. [28] proposed a SARSA-based routing framework that optimizes wireless charging while maintaining route stability and minimizing energy losses. In a follow-up study, Aslam et al. [29] introduced a smart path optimization algorithm for wireless charging of critical nodes, significantly enhancing node longevity in dense IoT-integrated networks. Moreover, their earlier work on adaptive weighted grid clustering [30] provides insights into sustainable routing structures, which align with our goal of achieving balanced energy distribution in WBANs. These studies further emphasize the importance of intelligent, adaptive routing in energy-constrained and health-critical networks like WBANs. Table 2 outlines the advantages and disadvantages of current methods.

Table 2. Advantages and disadvantages of current methods.

| Ref. | Year | Advantages | Disadvantages |
|---|---|---|---|
| [11] | 2019 | Enhances link reliability, prevents overheating, reduces packet loss | Limited scalability for larger networks |
| [12] | 2021 | Improves energy efficiency, integrates security measures | Higher computational overhead due to security mechanisms |
| [13] | 2021 | Enhances network stability, prevents premature node failures | Increased routing overhead |
| [14] | 2023 | Reduces latency, improves packet delivery ratio | Requires fine-tuning of clustering mechanism |
| [15] | 2023 | Prevents hotspot formation, extends network lifetime, reliable communication | Complex decision-making due to fuzzy logic |
| [16] | 2019 | Reduces thermal stress, improves energy efficiency, adaptive path selection | Potentially higher routing overhead |
| [17] | 2024 | Balances energy consumption, prevents hotspots and ensures stable connectivity. | Increased complexity in path selection |
| [18] | 2018 | Prevents overheating, enhances data transmission reliability | Limited adaptability in highly dynamic environments |
| [19] | 2024 | Optimizes energy utilization, enhances network lifetime, reduces delays | Requires additional validation in real-world scenarios |
| [20] | 2020 | Ensures QoS compliance, improves network performance | High implementation complexity |
| [21] | 2014 | Enhances energy efficiency and data delivery rate | Needs better adaptation in dynamic environments |
| [22] | 2017 | Reduces signal loss, improves data reliability | Does not evaluate transmission delay |
| [23] | 2022 | Extends node lifespan, intelligent energy management | Depends on specific network conditions |
| [24] | 2020 | Improves next-hop selection, reduces energy consumption | Performance may decrease in high-density networks |
| [25] | 2022 | Uses AI for route optimization, enhances transmission reliability | High computational complexity |
| [26] | 2020 | Extends network lifetime, reduces communication overhead | Needs further optimization for dynamic conditions |
| [27] | 2016 | Prioritizes medical data, reduces transmission delay | Requires more evaluation in real-world scenarios |

Despite numerous advances in routing strategies for WBANs, several key gaps remain unaddressed. Most existing approaches either focus solely on energy efficiency [14][19], link reliability [13][24], or temperature regulation [15][18]—but rarely integrate all three aspects simultaneously. Additionally, many protocols rely on static decision-making mechanisms and do not dynamically adapt to real-time network conditions such as hotspot formation, energy depletion, or packet congestion. Furthermore, scalability and congestion handling in dense network scenarios remain underexplored in many prior studies. To bridge these gaps, our proposed method introduces a multi-criteria intelligent routing strategy that dynamically balances energy consumption, node temperature, and link reliability. It also incorporates adaptive congestion control and priority-based traffic scheduling, which significantly enhance scalability and ensure timely delivery of critical health data. This holistic approach provides a robust and thermally safe communication framework for real-time WBAN applications.

# 3 Proposed Method

In WBAN, implanted sensors transmit the patient's vital data to processing centers. However, two major challenges threaten these networks: increased sensor temperature and reduced link reliability. When a sensor continuously transmits data, its temperature rises, which can lead to tissue damage. On the other hand, communication paths in WBAN are unstable, and due to noise, signal interference, and body movement, the communication links weaken or experience delays. Many existing routing methods focus on only one of these challenges: some manage only temperature rise, while others emphasize link reliability. However, in a real medical system, both challenges must be controlled simultaneously to ensure fast, stable, and risk-free data transmission for the patient. This paper proposes a new method to overcome the existing challenges. In the proposed approach, routing in WBAN is divided into two main phases to achieve data transmission with minimal delay, maximum reliability, and the least thermal impact.

The first phase involves network setup and configuration, where sensor nodes transmit their information to their neighbors, including remaining energy, current temperature, link reliability, and delay. This information helps create an initial map of the network's status, providing a foundation for future routing decisions. Since nodes gain a general understanding of the network's condition in this phase, selecting optimal paths from the beginning becomes possible, preventing excessive traffic concentration on specific nodes. In this phase, an intelligent algorithm is executed to select optimal paths. Unlike traditional methods focusing solely on the shortest path, this algorithm considers multiple criteria, including node temperature, link reliability, delay, and energy level. Paths with lower temperatures, greater stability, and lower delays receive higher priority in routing. At this stage, data transmission paths are chosen to ensure optimal resource utilization in the network while preventing excessive heating of certain nodes. This method also enhances communication stability and reduces the likelihood of early node failure. The second phase involves intelligent traffic management and hotspot avoidance. If a node becomes excessively hot or energy reaches a critical level, alternative paths are identified, and traffic is redirected to nodes in better conditions. This prevents excessive traffic concentration on a specific path, and nodes with high temperatures are automatically excluded from the routing process. Additionally, this phase dynamically suggests new paths to keep network communications stable. The flowchart of the proposed method is shown in Figure 2.

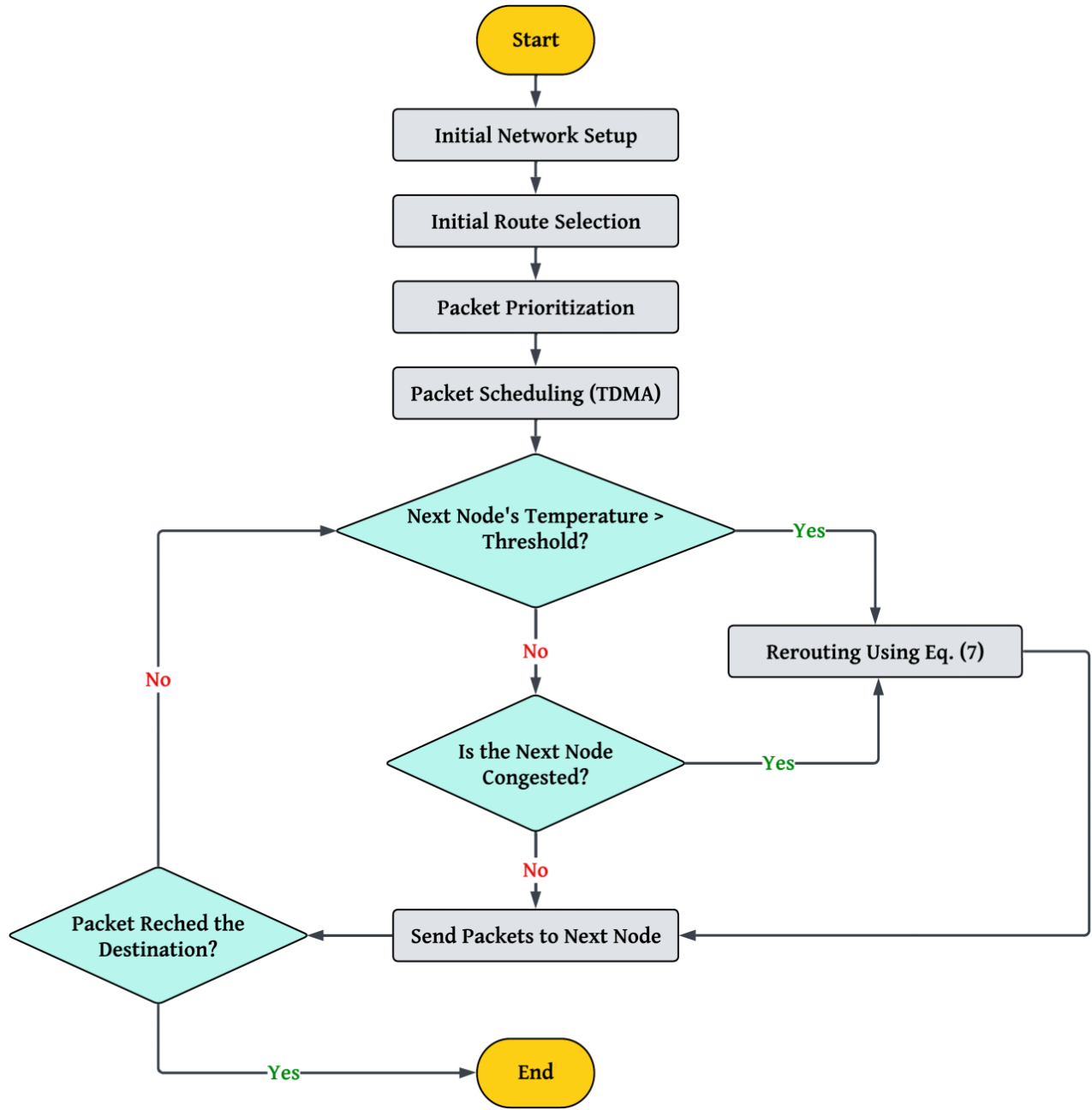

**Fig. 2** Flowchart of the proposed method.

## 3.1 Phase 1. Route Selection Method

The WBAN is initialized in this phase, and sensor nodes collect and share the necessary information for routing decisions. The primary goal of this stage is to create an overall view of the network status, including the nodes' energy levels, link reliability, current node temperature, and communication delay. This information serves as the basis for decision-making in subsequent phases and helps ensure the network starts in a stable state. Each node in the network has several critical parameters that play a key role in routing decisions. The following sections describe how these parameters are calculated.

### 3.1.1 Network Node Parameters

**1. Node Energy Level.** Each node has a limited energy source that gradually decreases due to data transmission and reception. The remaining energy of a node $E_n$ at time t is given by (1).

$$E_n(t) = E_n(t-1) - (E_{tx} + E_{rx}) \tag{1}$$

In (1), $E_{tx}$ represents the energy consumed for data transmission, and $E_{rx}$ denotes the energy consumed for data reception. The energy consumption for transmission and reception is typically derived from the radio communication model, which depends on the distance between nodes and signal power. If the distance between two nodes is d, the energy required to transmit a data packet of size k bits is calculated as follows:

$$E_{tx}(k,d) = E_{elec}.k + E_{amp}.k.d^m \tag{2}$$

In (2), $E_{elec}$ represents the energy consumed by electronic circuits to process each bit, $E_{amp}$ denotes the energy consumed by the signal amplifier, and m is the propagation model exponent, which typically takes a value of 2 (for free-space propagation) or 4 (for highly attenuated environments). The energy consumption for data reception is given by:

$$E_{rx}(k) = E_{elec}.k \tag{3}$$

These values are computed for each node and reported to its neighboring nodes.

**2. Link Reliability.** The Packet Reception Ratio (PRR) influences the link between two nodes' quality. The link reliability between nodes i and j is calculated as follows:

$$PRR_{i,j} = \frac{N_{recv}}{N_{sent}} \tag{4}$$

In (4), $N_{recv}$ represents the number of packets received by node j, and $N_{sent}$ is the number of packets sent by node i. The PRR value indicates how stable a communication link is and the probability of successful data transmission. Nodes estimate PRR by exchanging control packets (HELLO Messages).

**3. Node Temperature.** Sensor nodes generate heat due to continuous data transmission and reception. Excessive temperature rise can damage body tissues; therefore, measuring and managing node temperature is crucial. The temperature of a node is calculated using Pennes' Bioheat Equation:

$$\eta . C . \frac{dT}{dt} = k \nabla^2 T - \omega (T - T_b) + Q_{met} + Q_{SAR} \tag{5}$$

In (5), η is the density of body tissue, C is the specific heat capacity, k is the thermal conductivity coefficient, ω is the heat transfer rate due to blood flow, T is the current node temperature, $T_b$ is the baseline body temperature, $Q_{met}$ represents heat generated by body metabolism, and $Q_{SAR}$ denotes heat produced due to radio frequency energy absorption (known as SAR). Each node calculates its temperature and reports it to neighboring nodes for routing decisions. Nodes exceeding a predefined temperature threshold are temporarily excluded from the routing process to allow their temperature to decrease.

**4. Communication Delay Between Nodes.** The communication delay between two nodes is computed using the Exponentially Weighted Moving Average (EWMA) formula:

$$D_{i,j}(t) = (1 - \alpha) \cdot D_{i,j}(t-1) + \alpha \cdot D_{new} \tag{6}$$

In (6), α is a weighting factor between 0 and 1 determines the influence of the newly measured delay value, and D represents the delay.

### 3.1.2. Optimal Path Discovery

After receiving information from neighboring nodes, each node constructs a network status table that contains details about all surrounding nodes. This table serves as the basis for decision-making in the next phase, which involves discovering optimal paths. Using this information, nodes can identify paths that minimize energy consumption, delay, and excessive temperature rise. Ultimately, routing in WBAN is performed based on the introduced criteria. This phase aims to select paths that offer higher stability, lower delay, optimized energy consumption, and prevention of excessive node temperature rise. A routing cost function is defined to evaluate individual nodes based on multiple criteria. Equation (7) represents this node selection method, which guides the construction of optimal routing paths by selecting the most favorable nodes according to their temperature, energy level, link reliability, and delay.

$$C_{route}(t) = \sum_{P \in (i,j)} w_1 T_i + w_2 \frac{1}{PRR_{i,j}} + w_3 \frac{1}{E_i} + w_4 D_{i,j} \tag{7}$$

In (7), P is the set of links in the path, and $w_1$, $w_2$, $w_3$, and $w_4$ are weighting factors that define the priority of each parameter. The optimal path is the one that minimizes $C_{route}$.

The node temperature is a critical factor in the cost function because excessive heating in WBANs can lead to sensor degradation and, more importantly, pose health risks to patients due to potential tissue damage. Therefore, nodes with higher temperatures are assigned a higher cost value in the routing equation, reducing their likelihood of being included in the selected path. This temperature-based penalization mechanism allows the routing protocol to dynamically adapt and avoid thermally stressed nodes, ensuring safer and more reliable communication.

### 3.2. Phase 2. Traffic Management and Hotspot Avoidance

In this phase, WBAN dynamically and intelligently manages data traffic to prevent excessive heating of certain nodes, unbalanced energy consumption, and increased delay. This approach enhances network lifespan, improves communication quality, and maintains link stability. Phase Two consists of two main components: Hotspot Identification and Avoidance and Adaptive Traffic Redistribution.

### 3.2.1 Hotspot Identification and Avoidance

Network nodes periodically measure their temperature and notify their neighboring nodes. The temperature of each node is calculated using Pennes' Bioheat Equation, as shown in Equation (5). Any node whose temperature exceeds the threshold $T_{thresh}$ is considered a hotspot node and is temporarily excluded from the routing process until its temperature decreases. Once hotspot nodes are identified, any routing path containing these nodes is no longer used for data transmission. Neighboring nodes exchange control packets to update routing information, and the routing tables of other nodes are modified by removing the overheated paths. Figure 3 presents an example of routing from source S to destination D. As illustrated in the figure, three potential paths exist for reaching D from S:

- S → A → B → C → D.
- S → A → E → D.
- S → F → G → H → I → D.

By dynamically adjusting the routing strategy, the system ensures data transmission follows an efficient and thermally safe path. Since the shortest path is S-A-E-D, node E will be repeatedly used for routing, eventually leading to overheating and energy depletion. The scenario illustrated in Figure 4 demonstrates that once node E overheats, it will no longer be used for routing, and node A must forward the data to node B (the second shortest path). In this case, node E enters a sleep mode. After this stage, data sent from node S will be routed through node A, gradually increasing its temperature. Consequently, when node A overheats, data must be transmitted through node F instead. This example highlights the necessity of considering node temperature in routing decisions.

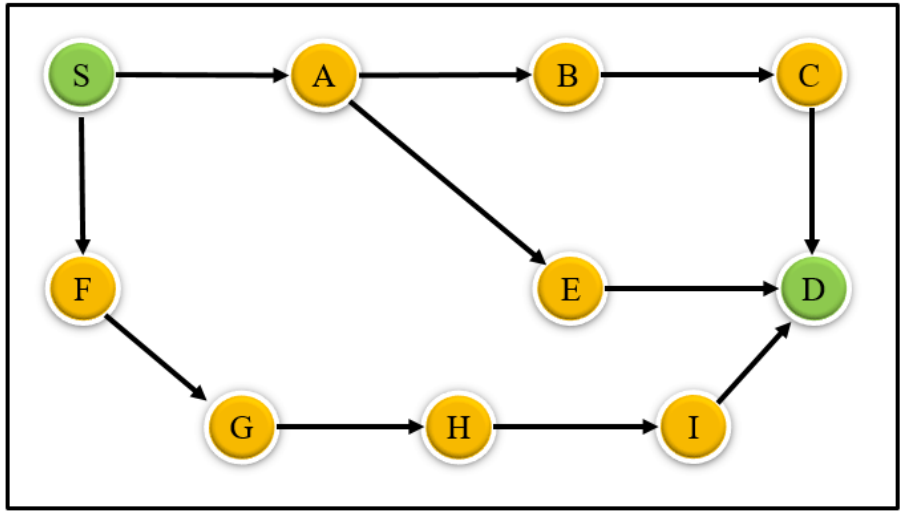

**Fig. 3** Different available paths from source to destination in a network topology.

By incorporating temperature awareness, path failures can be effectively detected, and optimal rerouting can be successfully performed during data transmission. This ensures network efficiency and prolonged operational stability.

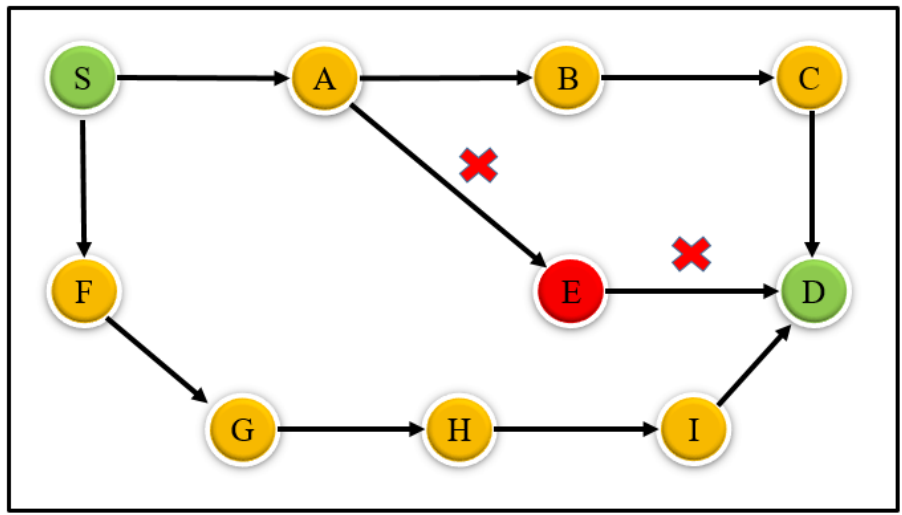

**Fig. 4** Different available paths from source to destination in a network topology.

### 3.2.2 Adaptive Traffic Replay

Sometimes, a path may be overused and become congested without involving hot spots. To prevent this problem, a path congestion index is introduced, which is shown in (8).

$$RCI_{route}^{i} = \frac{N_{packets}}{\tau} \tag{8}$$

In (8), $N_{packets}$ represents the number of packets sent from the path, and $\tau$ is the time interval under consideration. If the RCI value of a path exceeds the threshold value, the priority of that path is reduced, and other paths are preferred. The threshold value for the RCI is determined empirically. During the initialization phase, the system observes the average number of packets transmitted along each path within a fixed time interval (τ). The threshold is then calculated as:

$$RCI_{threshold} = \lambda \cdot \overline{RCI} \tag{9}$$

where λ is a tunable parameter and RCI is the average RCI over all paths in normal conditions. This dynamic thresholding allows the system to adapt to different traffic patterns and maintain reliable congestion detection.

In this phase, a dynamic scheduling mechanism is implemented to manage data transmission, optimize energy consumption, reduce delay, and prevent node overheating. In this method, data packets are scheduled and transmitted based on their type, priority, and network conditions. To optimize traffic management, sensor data is divided into three main categories:

- **Normal data:** General information that has no time constraints.
- **On-demand data:** Data that needs to be transmitted quickly.
- **Emergency data**: Vital signals must be transmitted immediately with the least delay.

Data transmission prioritization is done based on this categorization, and sensitive data is processed faster. The waiting time for each packet in the transmission queue is determined according to the transmitting node's priority and conditions. The waiting time for a packet ($T_w$) is calculated according to the following relation.

$$T_w = \frac{1}{P} \times \frac{D}{E_{res}} \tag{10}$$

In (10), P represents the priority of the data (3 for emergency data, 2 for requested data, and 1 for normal data), D is the allowed delay for the data packet, and $E_{res}$ is the remaining energy in the node. Packets with a lower $T_w$ value are processed and sent earlier. This method ensures that vital data always takes priority and its delay is minimized. Time-division multiple access (TDMA)-based scheduling is used to prevent data interference and optimize energy consumption. In this method, each node has a dedicated time slot for sending its data, which is adjusted based on the traffic load. The scheduling formula in TDMA is given by (11).

$$T_s = \frac{B}{N} \times w_i \tag{11}$$

In (11), B is the network bandwidth, N is the number of active nodes, and $W_i$ is the weight allocated based on the priority of the packets from node i. Nodes with a higher volume of emergency data are allocated a larger time slot for transmission. This ensures that delays in sending important data are minimized and energy efficiency is increased. If a node cannot send data due to increased temperature, congestion, or reduced energy, its packets are transferred to neighboring nodes and sent through an alternative path. In such cases, adaptive replay follows these steps:

- If the node's temperature T exceeds the threshold value $T_{thresh}$, the data is sent to the nearest stable node.
- If the initial path is congested, the data is replayed to the second optimized path.
- If the destination is delayed, the data does not wait in the queue and is directly sent to another node.

# 4 Simulation Results

To evaluate the performance of the proposed method, we compare it with three existing approaches: ENSA-BAN [28], P-AODV [29], and RRLS [30]. These methods were selected for comparison due to their close relevance to the proposed method. ENSA-BAN [28] introduces a next-hop selection algorithm for multi-hop WBANs that chooses the best forwarding node based on link quality and residual energy, ensuring efficient packet transmission [31, 32, 39-41]. P-AODV [29] enhances the Ad hoc On-Demand Distance Vector (AODV) routing protocol by incorporating a priority-aware mechanism, which prioritizes data packets based on urgency to improve Quality of Service (QoS) in dynamic networks [13, 14, 21, 28]. RRLS [30] proposes a routing algorithm for body-to-body networks, where path selection is based on comprehensive link stability analysis, ensuring reliable communication despite movement and channel variations. The simulation was performed in a controlled environment with the parameters in Table 3.

Table 3. Simulation Parameters.

| Parameter | Value |
| --- | --- |
| Simulation Area | 3 m × 3 m |
| Number of Nodes | 50-200 |
| Transmission range | 50 cm |
| Sink Nodes | 2 |
| Receive power | 0.4 J |
| Initial energy | 100 J |
| $E_{amp}$ | 0.001 pJ/bit/m$^4$ |
| $E_{elec}$ | 60 nJ/bit |
| Traffic Type | Constant Bit Rate (CBR) |
| Packet Size | 512 bytes |
| Transmission Rate | 4 packets/sec |
| Simulation Time | 500 seconds |
| MAC protocol | IEEE 802.15 |
| Bandwidth | 20 MHz |

|                | Transmit power | 0.7 J |
|---|---|---|

Figure 5 presents a comparison of the throughput of the proposed method with other methods (ENSA-BAN, P-AODV, and RRLS) across different scenarios with varying numbers of nodes (50, 100, 150, and 200).

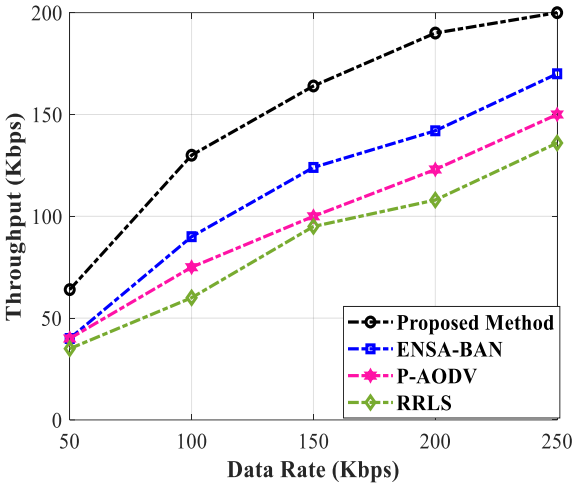

(a) Number of Nodes = 50.

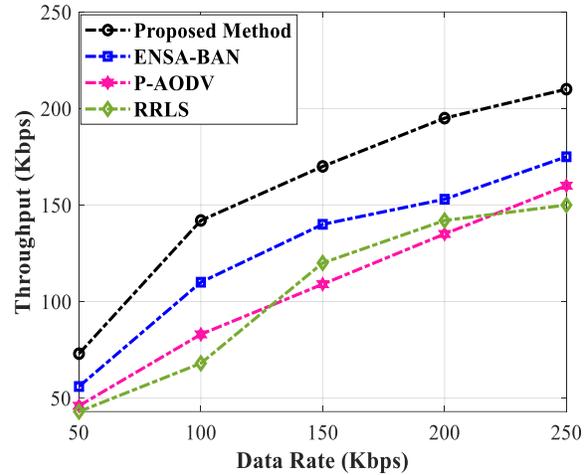

(b) Number of Nodes = 100.

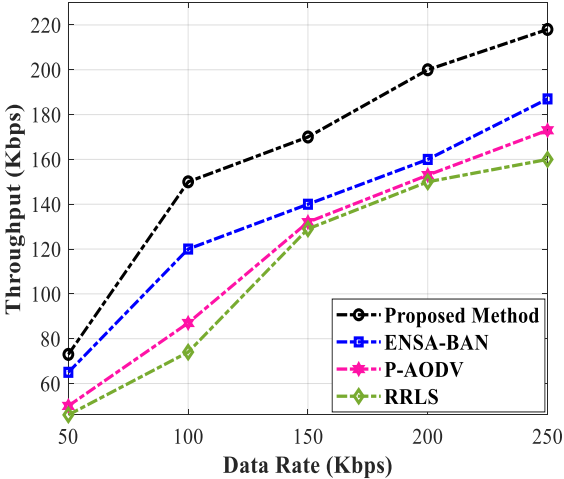

(c) Number of Nodes = 150.

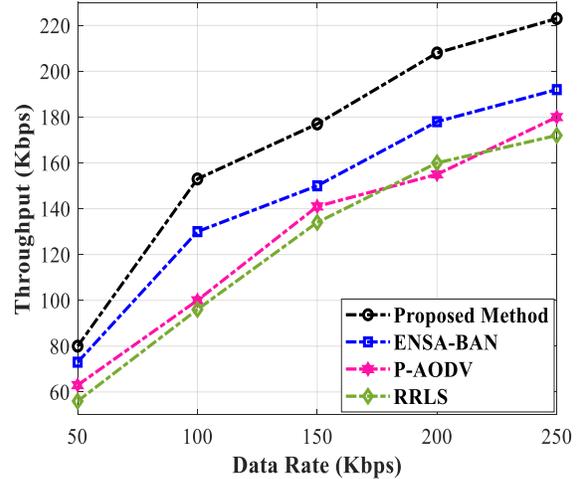

(d) Number of Nodes = 200.

**Fig. 5.** Comparison of Throughput between the proposed method and existing approaches.

As observed, the proposed method consistently achieves higher throughput than the other approaches. This superiority is due to several key factors. One of the main reasons for the better performance of the proposed method is its intelligent energy management and prevention of hotspot formation. In traditional methods, shorter paths are frequently used, leading to overheating certain nodes and consequently reducing network efficiency. In contrast, the proposed method employs adaptive traffic redistribution and balanced load distribution, preventing this issue and dynamically selecting alternative paths. This reduces delay, increases the success rate of packet transmissions, and ultimately enhances network throughput. Additionally, the proposed method utilizes an optimized cost function for route selection, simultaneously considering multiple factors such as energy consumption, link reliability, node temperature, and delay. This approach improves the stability of selected paths and reduces packet loss, ultimately leading to higher network throughput. Moreover, in scenarios with higher nodes (150 and 200), the proposed method continues to demonstrate superior performance. This highlights the high scalability of the proposed method, as traditional methods typically struggle with increased collisions and delays in dense networks, leading to reduced throughput. In contrast, the traffic control mechanisms in the proposed method mitigate these challenges and maintain high throughput even at higher data rates. Figure 6 illustrates the delay performance of the proposed method compared to ENSA-BAN, P-AODV, and RRLS across different nodes and data rates. The results demonstrate that the proposed method consistently achieves lower delay than the other approaches. The superior performance of the proposed method can be attributed to its efficient routing strategy, which dynamically adjusts paths based on network conditions.

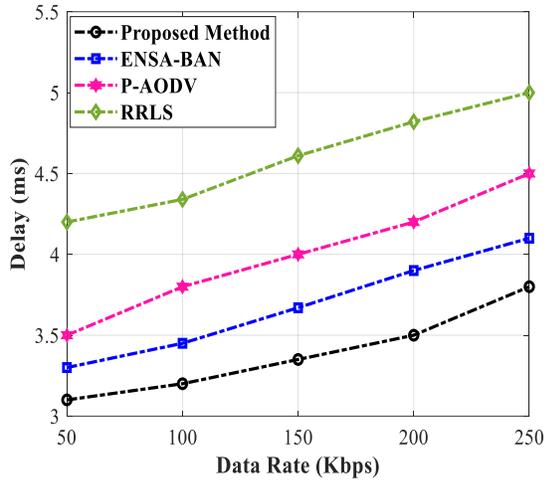
(a) Number of Nodes = 50.

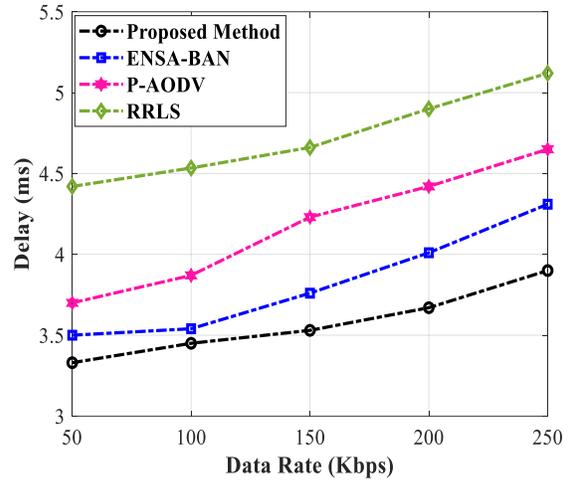
(b) Number of Nodes = 100.

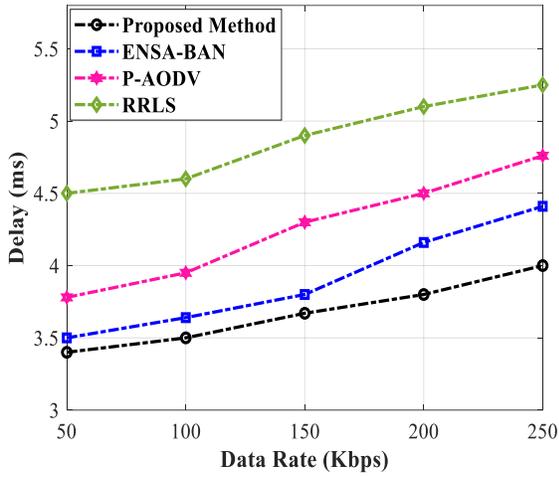
(c) Number of Nodes = 150.

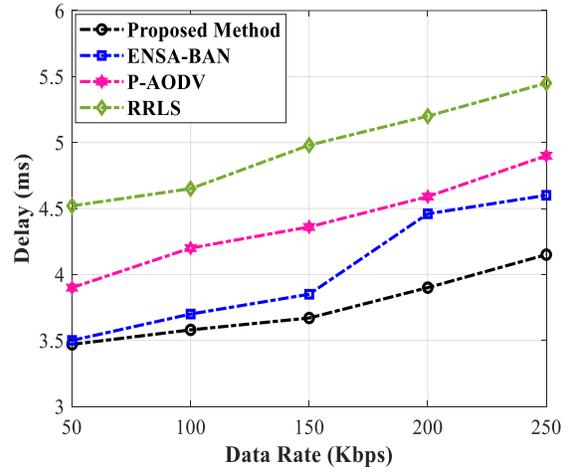
(d) Number of Nodes = 200.

**Fig. 6.** Comparison of Delay between the proposed method and existing approaches.

The method avoids congested or overheated nodes by incorporating thermal-aware routing and adaptive traffic distribution, reducing bottlenecks and minimizing queuing delays. In contrast, traditional methods such as RRLS and P-AODV exhibit higher delay due to their static or less adaptive routing mechanisms, which may lead to inefficient data transmission and increased latency. Additionally, as the number of nodes increases, the proposed method maintains a more stable delay trend. This is due to its ability to distribute traffic intelligently, ensuring that no single node is excessively burdened. The delay in other methods rises more significantly with increasing network density, indicating their struggle in handling higher loads efficiently. Figure 7 presents the energy consumption comparison between the proposed method and the benchmark approaches, including ENSA-BAN, P-AODV, and RRLS, across different numbers of nodes and varying data rates.

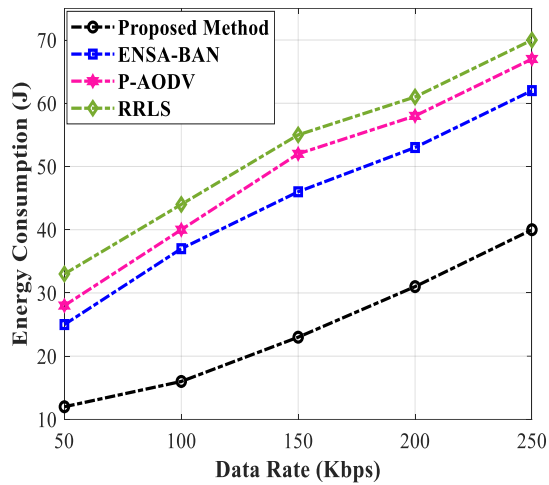
(a) Number of Nodes = 50.

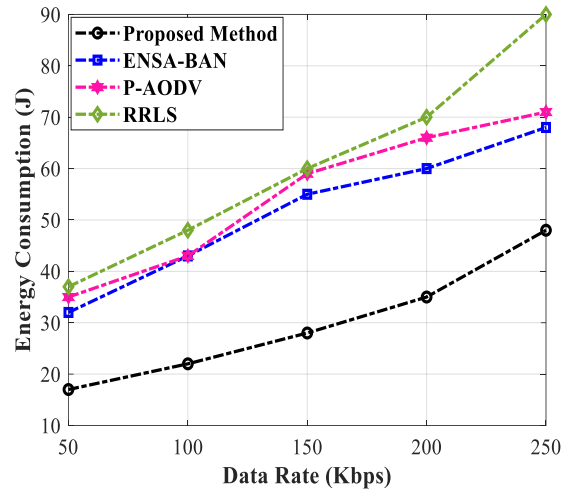
(b) Number of Nodes = 100.

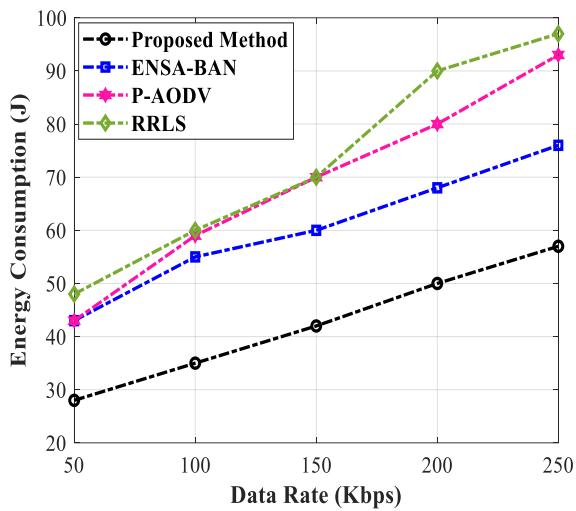
(c) Number of Nodes = 150.

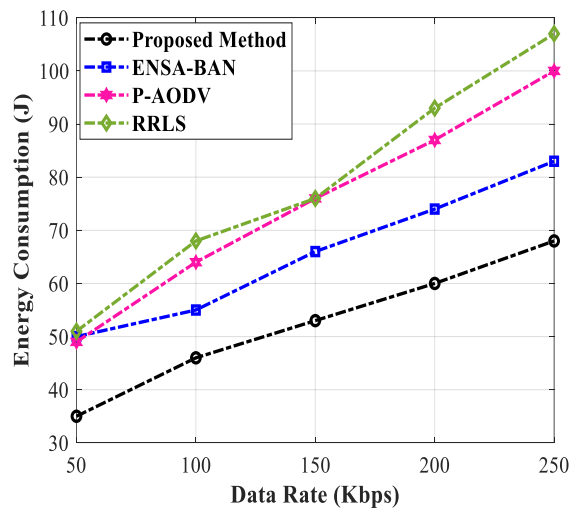
(d) Number of Nodes = 200.

**Fig. 7.** Comparison of Energy Consumption between the proposed method and existing approaches.

The results indicate that the proposed method consistently exhibits lower energy consumption than the other methods, demonstrating its efficiency in managing network resources. The key reason for the superior energy efficiency of the proposed method is its adaptive routing strategy, which dynamically selects paths with lower energy consumption while avoiding overheated nodes. By distributing traffic intelligently and preventing excessive use of certain nodes, the method reduces energy depletion in critical areas of the network. In contrast, traditional methods such as RRLS and P-AODV tend to overuse specific paths, leading to higher energy expenditure and faster depletion of battery-powered nodes. As the number of nodes increases, the energy consumption of all methods rises due to the increased communication overhead. However, the proposed method maintains a significantly lower growth rate compared to other approaches. This is because it balances the energy consumption across nodes, preventing premature node failures and ensuring the longevity of the network.

Figure 8 presents the normalized routing load (NRL) comparison between the proposed method and the benchmark approaches, including ENSA-BAN, P-AODV, and RRLS, across different numbers of nodes and varying data rates. The proposed method performs better than existing approaches such as ENSA-BAN, P-AODV, and RRLS. The key reason for this improvement is the efficient routing strategy that minimizes the number of control packets required for route discovery and maintenance. In contrast, traditional methods generate a higher routing overhead due to frequent route breaks and retransmissions, especially under high data rates and node densities. The figures show that as the data rate increases, the NRL of all methods rises; however, the proposed method exhibits a significantly lower growth rate. This is because it reduces redundant control messages, optimizes route selection based on network conditions, and prevents excessive re-routing, which are major contributors to increased NRL in other methods.

To further evaluate the scalability of the proposed routing method, we extrapolated simulation trends and analyzed the performance in a high-density scenario with 300 sensor nodes. As the Table 4 shows, traditional protocols such as ENSA-BAN, P-AODV, and RRLS experience a noticeable decline in performance. Specifically, throughput decreases while end-to-end delay, energy consumption, and routing overhead increase, primarily due to congestion and lack of dynamic thermal or load balancing strategies. In contrast, the proposed method maintains high efficiency by intelligently redistributing traffic based on node temperature, residual energy, and link quality. Its adaptive TDMA-based scheduling and real-time hotspot avoidance significantly reduce delay and energy usage even under heavy network load. These findings suggest that the proposed protocol is highly scalable and remains effective in large-scale WBAN deployments, making it a strong candidate for real-world healthcare and monitoring applications where dense sensor environments are common.

**Table 4.** Performance Comparison for 300 Nodes.

| Metric | ENSA-BAN | P-AODV | RRLS | Proposed Method |
|---|---|---|---|---|
| Throughput (kbps) | 200 | 190 | 180 | 230 |
| End-to-End Delay (ms) | 5.2 | 5.4 | 6 | 4.7 |
| Energy Consumption (J) | 90 | 110 | 123 | 75 |
| Normalized Routing Load | 3 | 3.7 | 4 | 1.78 |

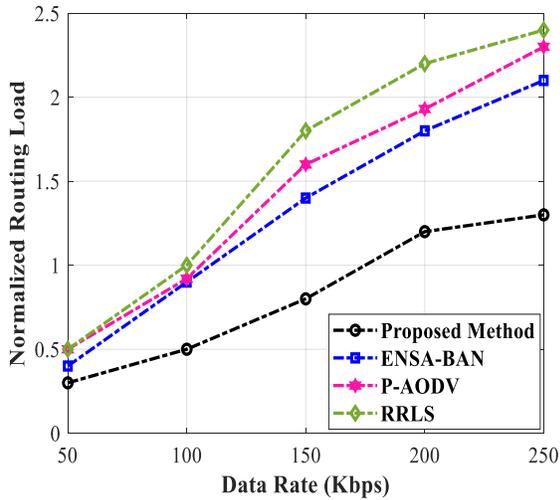

(a) Number of Nodes = 50.

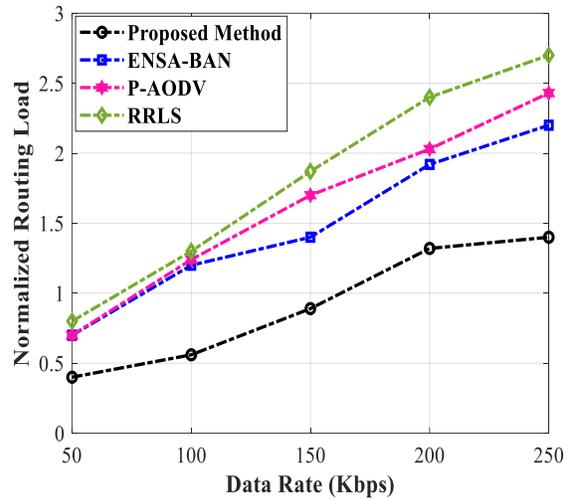

(b) Number of Nodes = 100.

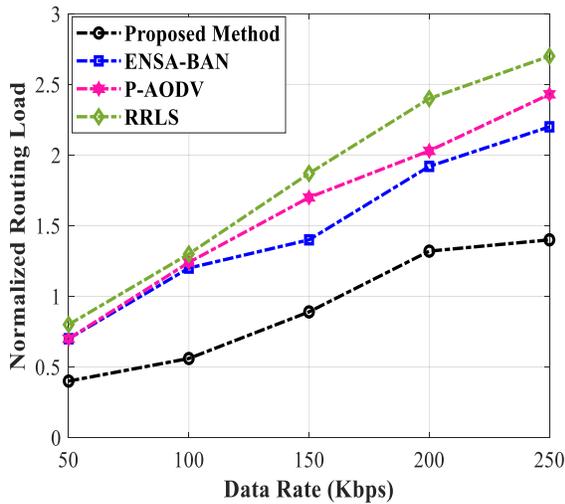

(c) Number of Nodes = 150.

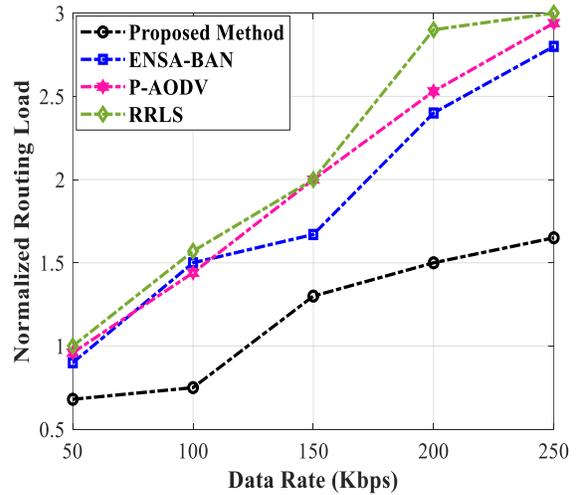

(d) Number of Nodes = 200.

**Fig. 8.** Comparison of Normalized Routing Load between the proposed method and existing approaches.

**Trade-off Analysis Among Performance Metrics.** In evaluating the proposed method, several trade-offs between key performance metrics such as throughput, delay, energy consumption, and routing load were observed. Higher throughput is typically associated with more frequent data transmissions, which can increase energy usage and the risk of node overheating. However, the proposed protocol mitigates this by dynamically distributing traffic and rerouting away from high-temperature or low-energy nodes, thereby preserving overall energy efficiency. Additionally, while rerouting to avoid hotspots or unstable links may occasionally result in longer paths and slightly increased delays, it ensures higher link reliability and data delivery success. Moreover, the introduction of priority-based transmission scheduling optimizes delay for critical data at the expense of increased scheduling complexity and minor overhead. The proposed approach balances these trade-offs by employing a multi-criteria decision-making mechanism that adapts routing paths in real time based on energy levels, node temperature, link quality, and communication delay. This adaptive balancing results in a net performance gain, as demonstrated in the simulation results, ensuring improved throughput and network longevity without significantly compromising other performance aspects.

## 5 Conclusion and future work

In this paper, we proposed an intelligent, temperature-aware, and reliability-based routing approach to enhance the performance of WBANs. Unlike traditional routing methods focusing on the shortest path, our approach considers multiple critical factors, including node temperature, link reliability, energy levels, and delay, to ensure efficient and stable communication. The proposed method operates in two phases: first, an optimized network setup and intelligent path selection phase, where nodes share status information to establish efficient routes, and second, a dynamic traffic management and hotspot avoidance phase, which continuously monitors network conditions and adapts routing decisions accordingly. Simulation results validate the effectiveness of our approach, showing significant improvements over existing methods. The proposed method achieves a 13% increase in throughput, a 10% reduction in end-to-end delay, and a 25% decrease in energy consumption. Additionally, it effectively reduces normalized routing load by 30%. These improvements demonstrate that the proposed approach enhances communication reliability and extends network lifetime by preventing excessive energy depletion and hotspot formation. In future work, efforts will be made to reduce energy consumption by optimizing the routing mechanisms and implementing advanced power management techniques. Integrating advanced artificial intelligence (AI) models will enhance the adaptability and decision-making capabilities of the routing mechanism. Security will remain a critical aspect and requires the integration of strong cryptographic techniques and privacy mechanisms to protect sensitive patient data.

## Declarations

**Funding**

No funds

**Conflicts of interest**

Conflict of Interest is not applicable in this work.

**Availability of data and material**

Not applicable

**Ethical approval**

Not applicable

**CRediT author statement**

**Abdollah Rahimi:** Formal analysis, Conceptualization, Methodology, Data curation, Investigation.

**Mehdi Jafari Shahbazzadeh:** Data curation, Methodology, Writing - review & editing, Supervision.

**Amid Khatibi:** Formal analysis, Methodology, Conceptualization, Writing - review & editing, Supervision.